\documentclass[aps,prl,preprint,longbibliography,superscriptaddress]{revtex4-2}
\usepackage{mathrsfs}
\usepackage{amsmath,gensymb}
\usepackage{amsfonts}
\usepackage{amssymb}
\usepackage{amsthm}
\usepackage{graphicx}
\usepackage{natbib}
\usepackage{color}
\usepackage{hyperref}
\usepackage{bm}
\usepackage[caption=false]{subfig}
\usepackage{verbatim}

\begin{document}
\title{Magnon-Cooparons in magnet-superconductor hybrids}

\author{Irina V. Bobkova}
\affiliation{Institute of Solid State Physics, Chernogolovka, 142432 Moscow region, Russia}
\affiliation{Moscow Institute of Physics and Technology, Dolgoprudny, 141700 Moscow region, Russia}
\affiliation{National Research University Higher School of Economics, 101000 Moscow, Russia}

\author{Alexander M. Bobkov}
\affiliation{Institute of Solid State Physics, Chernogolovka, 142432 Moscow region, Russia}
\affiliation{Moscow Institute of Physics and Technology, Dolgoprudny, 141700 Moscow region, Russia}

\author{Akashdeep Kamra}
\affiliation{Condensed Matter Physics Center (IFIMAC) and Departamento de F\'{i}sica Te\'{o}rica de la Materia Condensada, Universidad Aut\'{o}noma de Madrid, E-28049 Madrid, Spain}

\author{Wolfgang Belzig}
\affiliation{Fachbereich Physik, Universit{\"a}t Konstanz, D-78457 Konstanz, Germany \\ email: wolfgang.belzig@uni-konstanz.de}

\begin{abstract}
Generation and detection of spinful Cooper pairs in conventional superconductors has been intensely pursued by designing increasingly complex magnet-superconductor hybrids. Here, we demonstrate theoretically that magnons with nonzero wavenumbers universally induce a cloud of spinful triplet Cooper pairs around them in an adjacent conventional superconductor. The resulting composite quasiparticle, termed magnon-cooparon, consists of a spin flip in the magnet screened by a cloud of the spinful superfluid condensate. Thus, it inherits a large effective mass, which can be measured experimentally. Furthermore, we demonstrate that two magnetic wires deposited on a superconductor serve as a controllable magnonic directional coupler mediated by the nonlocal and composite nature of magnon-cooparons. Our analysis predicts a quasiparticle that enables generation, control, and use of spinful triplet Cooper pairs in the simplest magnet-superconductor heterostructures.
\end{abstract}

\maketitle

\section{Introduction}

The widely available and used conventional superconductors consist of spin-singlet Cooper pairs which are devoid of a net spin. Unconventional superconductors, in contrast, host qualitatively distinct phenomena and Cooper pair properties~\cite{Sigrist1991}. Their limited experimental availability, however, has driven the scientific community to try and engineer heterostructures comprising conventional superconductors into effectively unconventional ones~\cite{Bergeret2005,Eschrig2008,Buzdin2005,Linder2015,Bergeret2018}, e.g., in achieving Majorana bound states~\cite{Fu2008}. In particular, the highly desired spinful spin-triplet Cooper pairs can be generated from a conventional superconductor if the latter interacts with two or more non-collinear magnetic moments~\cite{Bergeret2005,Eschrig2008,Buzdin2005,Linder2015,Bergeret2018}. With this design principle, a wide range of magnet-superconductor hybrids with multiple magnetic layers to generate and detect spinful Cooper pairs have been investigated~\cite{Keizer2006,Khaire2010,Jeon2018,Jeon2020}. The challenge of detecting a spin or its flow directly has resulted in the need for increasingly complex magnet-superconductor hybrids rendering their direct detection a highly demanding, debated, and pursued goal~\cite{Bell2008,Jeon2018B,Yao2018,Mueller2021}.

The ambition is to go beyond detection, and towards exploiting the fascinating physics of these unconventional Cooper pairs for phenomena that are otherwise out of reach~\cite{Eschrig2015}. Noncollinear ground states of magnets~\cite{Robinson2010,Chiodi2013} and spin-orbit coupling~\cite{Johnsen2019,Ruano2020} have been exploited in generating equilibrium spinful Cooper pairs. These have allowed a control over static properties, such as magnetic anisotropy~\cite{Johnsen2019,Ruano2020} or superconducting critical temperature~\cite{Pugach2017}, of various superconductor-magnet hybrids. Nevertheless, on-demand steering and movement of spinful Cooper pairs is highly desired and has remained an outstanding challenge. For example, a directed flow of the spinful Cooper pairs could be used for delivering nondissipative spin transfer torques and magnetic switching~\cite{Bobkova2018,Silaev2020B,Linder2011,Waintal2002,Kulagina2014,Halterman2016}. Such goals face a similar challenge that even when a complex heterostructure generates spinful Cooper pairs, it becomes difficult to steer them. An injected charge current predominantly converts into conventional spinless supercurrent~\cite{Bobkova2018}. Due to such reasons, several advantages of magnet-superconductor heterostructures realized in various concepts and devices are still dominated by the quasiparticle properties~\cite{Machon2013,Ozaeta2014,Kolenda2016,Bergeret2018,Kato2019,Jeon2020B,Amundsenarxiv,Bobkova2021}. The exploiting of spinful Cooper pairs for exciting physics and devices has been impeded by the complex hybrids needed to generate them and the difficulty of steering them.

\begin{figure}[tbh]
	\begin{center}
		\includegraphics[width=100mm]{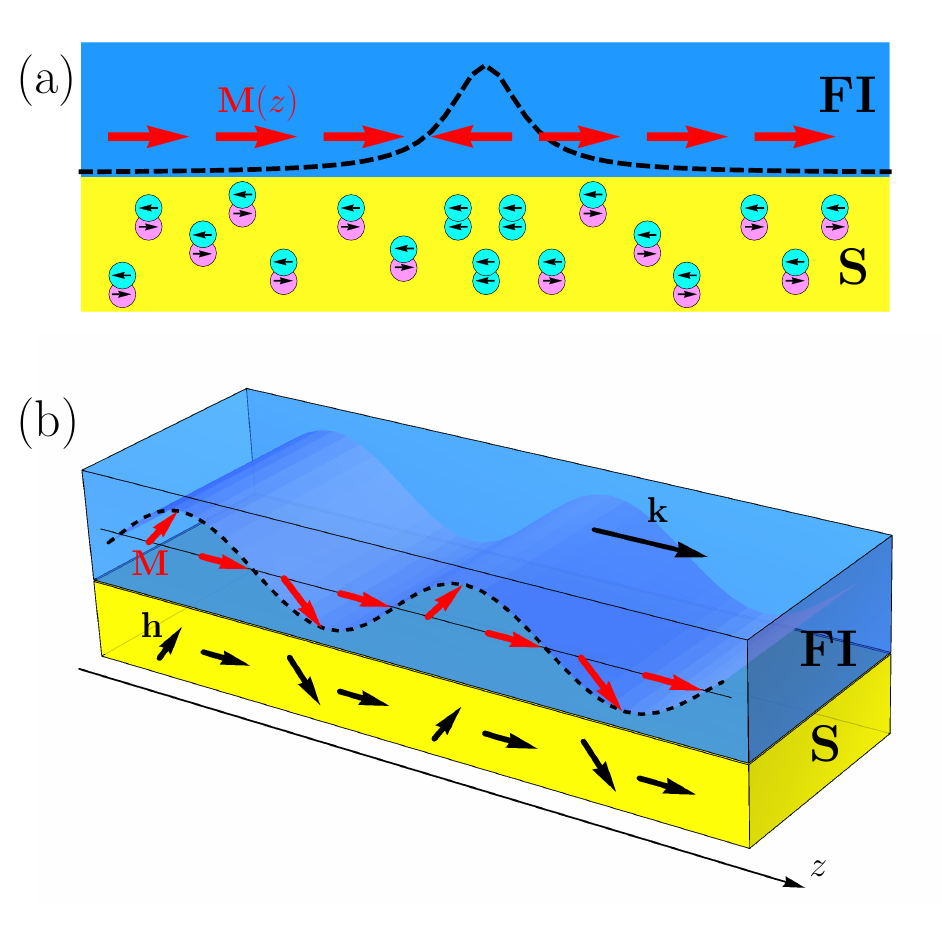}
		\caption{\textbf{Schematic depiction of the system under investigation and the magnon-cooparon quasiparticle.} (a) A localized spin-flip or magnon wave packet induces a surrounding cloud of spinful triplet Cooper pairs in an adjacent conventional spin-singlet superconductor. The spatially varying magnetization or spin profile (depicted via dashed line) associated with the excitation induces spinful condensate, that screens the magnon spin, in the otherwise spinless superconductor. The resulting quasiparticle, termed magnon-cooparon, bears a smaller spin and larger effective mass. Note that we depict magnetic moments (which point opposite to spins on account of the negative gyromagnetic ratio) in the FI, while depicting electronic spins in S. (b) A spin wave with wavenumber $k$ propagates in an in-plane direction. The associated noncollinear magnetization profile (red arrows) induces an analogous spatially varying exchange field (black arrows) in the adjacent superconductor, resulting in spinful triplet condensate.} \label{Fig1}
	\end{center}
\end{figure}

In this work, we uncover a ubiquitous existence and control of spinful Cooper pairs in the simplest magnet-superconductor hybrid - a bilayer - that has escaped attention thus far. We find that a magnon, the quasiparticle of spin waves in a magnet, with nonzero wavevector induces a cloud of spinful Cooper pairs in the adjacent superconductor [Fig.~\ref{Fig1}(a)]. This accompanying cloud screens the magnon spin giving rise to a composite heavy quasiparticle with an enhanced effective mass, which is termed `magnon-cooparon' due to its similarity to the polaron quasiparticle as discussed below. This induction of spinful Cooper pairs in a conventional superconductor is caused by the noncollinear magnetization profile of a spin wave with finite wavevector [Fig.~\ref{Fig1}(b)], an effect not seen when considering ferromagnetic resonance of the uniform magnon mode. Furthermore, we demonstrate theoretically that magnon-cooparons enable a magnonic directional coupler~\cite{Sadovnikov2015,Wang2018} composed of two separate ferromagnetic wires with coupling lengths shorter than previously feasible thereby allowing smaller devices. Thus, it enables a valuable application in magnon-based logic and circuits~\cite{Chumakarxiv,Pirro2021}. The magnon-cooparon is reminiscent of the fermionic polaron quasiparticle created by screening of an electron by a phonon cloud~\cite{Froehlich1954}, although the magnon-cooparon is a bosonic excitation. Considering the gradual discovery of polaron and its variants in a wide range of phenomena~\cite{Froehlich1954,Chuev1996}, we expect magnon-cooparon to find a similar important role in a broad range of magnet-superconductor hybrids. This concept can also significantly expand the range of reported effects related to the mutual influence of superconductivity and magnons~\cite{Skadsem2011,Simensen2021,Gusev2021,Jeon2018,Jeon2020,Bell2008,Jeon2018B,Yao2018,Kato2019,Jeon2020B,Amundsenarxiv,Golovchanskiy2020,Golovchanskiy2022,Ojajarviarxiv,Dobrovolskiy2019,Rogdakis2019,Johnsen2021}.

\section{Results}

\subsection{Emergence and effective mass of magnon-cooparons}

We consider a bilayer as depicted in Fig.~\ref{Fig1}(b), in which a ferromagnetic insulator FI (e.g., yttrium iron garnet) is interfaced with a conventional spin-singlet s-wave superconductor S (e.g., Nb). The two layers with thicknesses $d_{\mathrm{FI}}$ and $d_{\mathrm{S}}$ ($\ll \xi_{\mathrm{S}}$, the superconducting coherence length) are considered thin such that physical properties vary only in the in-plane direction. In its ground state, the FI is assumed to be magnetized along the z direction. For S/FI structures the effective induced  exchange field in the superconductor is well-documented experimentally by measurements of the spin-split DOS \cite{Bergeret2018}. At the same time for S/FM heterostructures, where FM means a ferromagnetic metal, the well-pronounced homogeneous spin-split DOS was not reported. The physical reason for this can be related to the leakage of Cooper pairs into the ferromagnet and, consequently, much stronger suppression of superconductivity at S/FM interfaces. Therefore, we expect that the renormalization of the magnon spin and stiffness by the cloud of triplet pairs, generated in the superconductor should be smaller in S/FM structures.
From the other hand, in S/FM heterostructures there is a proximity effect, that is a penetration of Cooper pairs into the ferromagnet. In principle, in this case the cloud of triplet pairs, screening the magnon, could be generated directly in the ferromagnetic metal. We expect that qualitative physics of the renormalization should be similar.

We wish to examine wavevector-resolved excitations of the hybrid, and thus obtain the complete information needed for examining arbitrary wavepackets generated by a given experimental method. To this end, we assume existence of a spin wave with wavevector $k \mathbf{e}_z$ in the FI [Fig.~\ref{Fig1}(b)] such that the magnetization unit vector $\mathbf{m}(\mathbf{r},t) = \mathbf{m}_0 + \delta \mathbf{m}(\mathbf{r},t)$ consists of the equilibrium part $\mathbf{m}_0 = \mathbf{e}_z$ and the excitation part $\delta \mathbf{m}(\mathbf{r},t) = \delta m \left[ \cos(kz + \omega t) \mathbf{e}_x + \sin(kz + \omega t) \mathbf{e}_y \right] \exp(- \kappa t)$. While we consider an excitation with wavevector along $\mathbf{e}_z$, our analysis is general and valid for any in-plane wavevector. The magnetization dynamics is described within the Landau-Lifshitz-Gilbert framework as
\begin{align}\label{eq:llg}
\dot{\mathbf{m}} & =  - \gamma \left( \mathbf{m} \times \mathbf{H}_{\mathrm{eff}} \right) + \alpha \left(\mathbf{m} \times \dot{\mathbf{m}} \right) + \tilde{J} \left(\mathbf{m} \times \mathbf{s}\right),
\end{align}
where $- \gamma$ with $\gamma>0$ is the FI gyromagnetic ratio, $\alpha$ is the Gilbert damping parameter, $\mathbf{H}_{\mathrm{eff}}$ is the effective magnetic field in the FI, and $\tilde{J} \equiv J/d_{\mathrm{FI}}$ with $J$ parameterizing interfacial exchange interaction between FI and S. The last term on the right hand side of Eq.~\eqref{eq:llg} accounts for the spin torque exerted on the magnetization by the spin density $\mathbf{s}$ it induces in S~\cite{Ralph2008}. Expressing $\mathbf{s} = s_0 \mathbf{m}_0 + \delta s_{\parallel} \delta \mathbf{m} + \delta s_{\perp} (\delta \mathbf{m} \times \mathbf{m}_0)$ and substituting the expressions for $\mathbf{s}$ and $\mathbf{m}$ in Eq.~\eqref{eq:llg} above, we obtain
\begin{align}
 \omega & = D_{\mathrm{m}} k^2 + \gamma K + \tilde{J} \left( \delta s_{\parallel} - s_0 \right), \label{eq:omega} \\
\kappa & = \alpha \omega - \tilde{J} \delta s_{\perp}, \label{eq:kappa}
\end{align}
where $D_{\mathrm{m}}$ is the FI spin wave stiffness and $K$ parameterizes a uniaxial anisotropy. Thus, the spin density $\mathbf{s}$ induced in the S may renormalize both the excitation frequency and its lifetime.

We now evaluate the induced spin density $\mathbf{s}$ treating S using the quasiclassical Green's functions framework~\cite{Belzig1999,Bergeret2018,Buzdin2005}. Working in the dirty limit, we need to solve the Usadel equation for the $8\times8$ matrix Green's function $\check{g}$ in spin, particle-hole, and Keldysh spaces:
\begin{align}\label{eq:usadel}
\mathrm{i} D \mathbf{\nabla} \left( \check{g} \otimes \mathbf{\nabla} \check{g} \right) & = \left[\epsilon \hat{\mathbb{I}} \hat{\tau}_z \hat{\mathbb{I}} - \mathbf{h} \cdot \hat{\pmb{\sigma}} \hat{\tau}_z \hat{\mathbb{I}} + \mathrm{i} \Delta \hat{\mathbb{I}} \hat{\tau}_y \hat{\mathbb{I}}, \check{g} \right]_{\otimes},
\end{align}
where $\hat{\mathbb{I}}$ is the $2\times2$ identity matrix, outer-product between the $2\times2$ matrices (decorated by overhead $\hat{}~$) in obtaining an $8 \times 8$ matrix (identified via an overhead $\check{}~$) is implied, and we set $\hbar = 1$ throughout this Letter. Further, working in the mixed ($\epsilon,t$) representation, we employ the notation $[A,B]_{\otimes} \equiv A \otimes B - B \otimes A$ with $A \otimes B \equiv \exp [(\mathrm{i}/2) (\partial_{\epsilon_1} \partial_{t_2} - \partial_{\epsilon_2} \partial_{t_1})] ~ A(\epsilon_1,t_1) B(\epsilon_2,t_2)|_{\epsilon_1 = \epsilon_2 = \epsilon, t_1 = t_2 = t} $. $\hat{\tau}_{x,y,z}$ and $\hat{\sigma}_{x,y,z}$ are Pauli matrices in particle-hole and spin spaces, respectively. A real $\Delta$  accounts for the intrinsic conventional spin-singlet order parameter of the superconductor. The term $\mathbf{h} \cdot \hat{\pmb{\sigma}}$ accounts for the exchange field induced by the adjacent FI layer~\cite{Tokuyasu1988,Hao1991,Buzdin2005,Moodera2007,Bergeret2018,Kamra2018,Cottet2009,Eschrig2015bc}. The exchange field bears units of energy, and corresponds to
the spin-splitting it causes. For the magnetization profile associated with the excitation under consideration, we obtain
\begin{align}\label{eq:h}
\mathbf{h} & = h_0 \mathbf{e}_z + \delta h \left[ \cos(kz + \omega t) \mathbf{e}_x + \sin(kz + \omega t) \mathbf{e}_y \right],
\end{align}
where $h_0$ and $\delta h$ respectively capture the static and dynamic components of the induced exchange field $\mathbf{h} = J M_{\mathrm{s}} \mathbf{m}/ 2 \gamma d_{\mathrm{S}}$, with $M_{\mathrm{s}}$ the FI saturation magnetization. The contribution of the superconductor dynamics to the excitation under consideration can be evaluated by solving Eqs.~\eqref{eq:usadel} and \eqref{eq:h} up to the first order in $\delta h$.
The desired spin density in the superconductor is evaluated as~\cite{Bergeret2018}
\begin{align}\label{eq:s}
\mathbf{s} & = - \frac{N_{\mathrm{F}}}{16} \int d\epsilon ~ \mathrm{Tr}_4 \left[ \left(\hat{\pmb{\sigma}} \hat{\tau}_z \right)  \breve{g}^{\mathrm{K}} \right],
\end{align}
where $\mathrm{Tr}_4$ denotes trace over a $4\times4$ matrix (decorated by an overhead $\breve{}~$), $\breve{g}^{\mathrm{K}}$ is the $4 \times 4$ Keldysh component of the full $8 \times 8$ Green's function $\check{g}$, and $N_{\mathrm{F}}$ is the normal state density of states at the Fermi level in S.

\begin{figure}[tbh]
	\begin{center}
		\includegraphics[width=100mm]{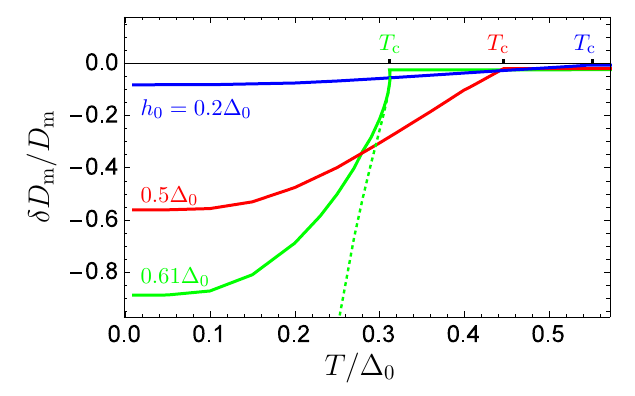}
		\caption{\textbf{Magnon-cooparon effective mass variation.} Relative change in the spin stiffness $\delta D_{\mathrm{m}} / D_{\mathrm{m}}$ as a function of temperature $T$ for different values of the static exchange field $h_0$ induced in S. A reduction of the spin stiffness signifies an increased effective mass of the composite excitation. $T_{\mathrm{c}}$ is the critical temperature of the superconductor. $\Delta_0$ is the superconducting gap when $T = h_0 = 0$. The dashed line plots the analytic result [Eq.~\eqref{eq:deltadm}] obtained in the limit $T \to T_{\mathrm{c}}$.} \label{Fig2}
	\end{center}
\end{figure}

Following the method outlined above and detailed in Supplementary Note 1, we obtain analytic expressions for $\mathbf{s}$ [Eq.~\eqref{eq:s}], and thus, the excitation frequency [Eq.~\eqref{eq:omega}] and lifetime [Eq.~\eqref{eq:kappa}]. These unwieldy expressions simplify considerably in the adiabatic limit of $\omega \ll T$, that we discuss first. Further, quasiparticles are found to not play an important role in this limit leaving the focus on the superfluid condensate. We find $\delta s_{\perp} \to 0$ in this limit such that the excitation decay rate [Eq.~\eqref{eq:kappa}] is not influenced by S. $\delta s_{\parallel} - s_0$ is found to scale as $\sim k^2$ in this limit, such that the excitation frequency becomes $\omega = \tilde{D}_{\mathrm{m}} k^2 + \gamma K$ with $\tilde{D}_{\mathrm{m}} \equiv D_{\mathrm{m}} + \delta D_{\mathrm{m}}$ and
\begin{align}\label{eq:deltadm}
\delta D_{\mathrm{m}} & = - \frac{\pi N_{\mathrm{F}} \gamma d_{\mathrm{S}} D \Delta^2}{16 T_{\mathrm{c}} d_{\mathrm{FI}} M_{\mathrm{s}}} \left[ \frac{1}{x}\tanh x - \frac{1}{\cosh^2x} \right],
\end{align}
where $x=h_0/2T_{\mathrm{c}}$, and $T_{\mathrm{c}}$ is the superconducting critical temperature taking into account the static exchange field $h_0$. In obtaining Eq.~\eqref{eq:deltadm}, we further worked in the limit $|T - T_{\mathrm{c}}| \ll T_{\mathrm{c}}$. The same stiffness renormalization [Eq.~\eqref{eq:deltadm}] is obtained from purely energy considerations within the Ginzburg-Landau framework~\cite{Silaevarxiv}. The effective mass $m_{\mathrm{eff}}$ of the composite quasiparticle is obtained as $m_{\mathrm{eff}} = 1/2\tilde{D}_{\mathrm{m}} = 1/(2 D_{\mathrm{m}} + 2 \delta D_{\mathrm{m}})$. Since $\delta D_{\mathrm{m}} < 0$ [Eq.~\eqref{eq:deltadm}], the effective mass of the composite quasiparticle is enhanced as compared to that of a magnon. Numerically evaluated $\delta D_{\mathrm{m}}$, without making the adiabatic approximation, plotted in Fig.~\ref{Fig2} versus temperature further shows the direct role of the superconducting condensate and suggests temperature as a handle to control the quasiparticle effective mass. The material parameters assumed in Fig.~\ref{Fig2} are detailed further below, together with the discussion on experimental detection.

Thus, this composite quasiparticle shares some similarities with the polaron~\cite{Froehlich1954}. The latter, predicted almost a century ago~\cite{Landau1933,Pekar1946} and having found numerous applications throughout condensed matter physics~\cite{Froehlich1954,Chuev1996}, is formed when an electron is screened by the phonon cloud leading to a heavy fermionic excitation. The quasiparticle under consideration is a bosonic magnon spin being screened by a superconducting condensate. Due to this similarity (and yet many distinctions) with the polaron, we term this spin flip surrounded by a spinful Cooper pairs cloud [Fig.~\ref{Fig1}(a)] magnon-cooparon. Physically, a finite wavenumber $k$ is needed to create spinful Cooper pairs, in an otherwise spinless conventional superconductor, via a noncollinear exchange field [Fig.~\ref{Fig1}(b)]. This explains the $\sim k^2$ dependence of the frequency renormalization, as well as the previous studies investigating uniform ($k = 0$) magnon modes not encountering the magnon-cooparon.

Finally, going beyond the adiabatic approximation $\omega \ll T$, we find a nonzero renormalization of the $k = 0$ mode frequency and the decay rate $\sim k^2$, as detailed in the Supplementary Note 2. Similar effects are expected based on spin pumping into a normal metal or quasiparticles in a superconductor~\cite{Tserkovnyak2002,Silaev2020,Kato2019}. Specifically, when quasiparticle spin relaxation is disregarded, the increase in decay rate requires a spin sink, which may be provided by the noncollinear magnetic moment in a second magnet~\cite{Tserkovnyak2005}. In our case, a spatially distinct part of the same magnet provides the noncollinear spin absorption channel, thereby short-circuiting the spin wave. Hence, in this case too, the unique dynamic noncollinearity of the finite-$k$ spin wave results in novel effects.

\subsection{Spin of magnon-cooparons}

The cloud of spinful Cooper pairs screening the magnon spin that increases its effective mass further implies that (i) the total spin of magnon-cooparon is reduced from 1, and (ii) a magnon spin current $j_{\mathrm{m}} \mathbf{e}_z$ in FI is accompanied by a superfluid spin current $j_{\mathrm{S}} \mathbf{e}_z$ in S. We now address these effects and ascertain the net spin of the magnon-cooparon.

Since the dc spin current $j_{\mathrm{m}}$ accompanying a spin wave or magnon scales as $\delta m^2$, we anticipate $j_{\mathrm{S}}$ to scale as $\delta h^2$, confirming this via a rigorous calculation detailed in the Supplementary Note 3. As a result, we now need to solve Eq.~\eqref{eq:usadel} for the matrix Green's function up to the second order in $\delta h$. Since this is a more demanding calculation than that carried out above, we restrict ourselves to the adiabatic approximation $\omega \ll T$ and the limit $|T - T_{\mathrm{c}}| \ll T_{\mathrm{c}}$ in the rest of our analysis. The spin current flowing along the direction of magnon propagation ($\mathbf{e}_z$ here) in S is then obtained as~\cite{Bergeret2018}
\begin{align}\label{eq:jS}
\mathbf{j}_{\mathrm{S}} \mathbf{e}_z & = \frac{N_{\mathrm{F}} D}{16} \int d\epsilon ~ \mathrm{Tr}_4 \left[ \left( \hat{\pmb{\sigma}} \hat{\mathbb{I}}  \right) \left( \check{g} \partial_z \check{g}\right)^{\mathrm{K}} \right] ~\mathbf{e}_z,
\end{align}
where the direction of $\mathbf{j}_{\mathrm{S}}$ pertains to the spin space. On explicit evaluation shown in the Supplementary Note 3, $\mathbf{j}_{\mathrm{S}}$ is found to bear only a z component, as can be expected from its screening the magnon spin, which itself bears only a z component. The total spin current may thus be expressed as
\begin{align}\label{eq:magcoopspin}
\mathbf{j}_{\mathrm{m}} + \mathbf{j}_{\mathrm{S}} = S v_{k} n_{k} ~ \mathbf{e}_z \equiv \left(1 + \frac{j_{\mathrm{Sz}}}{j_{\mathrm{m}}} \right) v_{k} n_{k} ~ \mathbf{e}_z,
\end{align}
where $v_k = 2 \tilde{D}_{\mathrm{m}} k$ is the magnon-cooparon group velocity, $n_k$ is the number of excitations, and $S$ becomes its net spin evaluated via
\begin{align}\label{eq:spincurrratio}
\frac{j_{\mathrm{Sz}}}{j_{\mathrm{m}}} & = - \frac{8 N_{\mathrm{F}} D \gamma}{ \tilde{D}_{\mathrm{m}} M_{\mathrm{s}}}~ \sum_{\omega_n > 0} \frac{ \pi T_{\mathrm{c}} \Delta^2 h_0^2 ~ \omega_n^2}{\left( \omega_n^2 + h_0^2 \right)^2 \left(2 \omega_n + D k^2 \right)^2},
\end{align}
where $\omega_n$ are the fermionic Matsubara frequencies. Since $j_{\mathrm{Sz}}/j_{\mathrm{m}} < 0$, the net spin of the magnon-cooparon is reduced from 1 as per our expectation from the screening. Equation \eqref{eq:spincurrratio} shows that the dynamical induction of spinful Cooper pairs always causes screening, and thus, a reduction in the excitation net spin. Further, similar to the relative change in the spin stiffness (Fig.~\ref{Fig2}), the spin reduction $|j_{\mathrm{Sz}}/j_{\mathrm{m}}| \lesssim 1$ for typical material parameters, as estimated further below.


\begin{figure}[tb]
	\begin{center}
		\includegraphics[width=110mm]{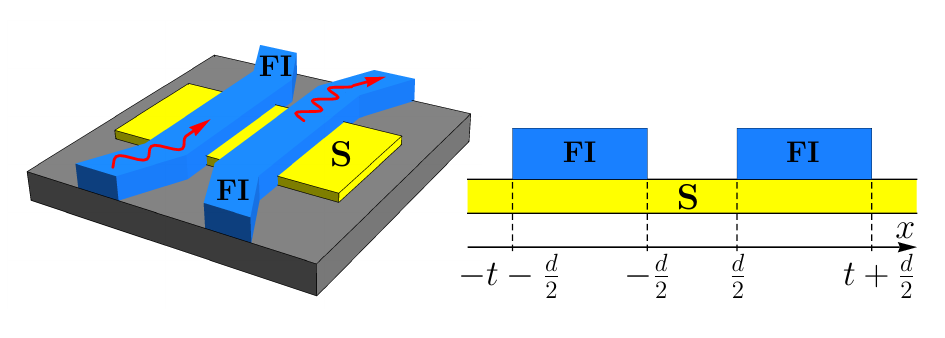}
		\caption{\textbf{Schematic depiction of a magnonic directional coupler based on magnon-cooparons}. A spin wave propagating through one FI wire is controllably transferred to the second FI wire.} \label{Fig3}
	\end{center}
\end{figure}

\subsection{Superfluid-mediated magnonic directional coupler}

Since the Cooper pairs cloud comprising a magnon-cooparon extends over a length $\sim \xi_{\mathrm{S}}$, it enables transfer of energy from a spin wave in one FI wire to another, nonlocally (Fig.~\ref{Fig3}). Thus, two FI wires deposited on a conventional superconductor within $\sim \xi_{\mathrm{S}}$ from each other act as a magnonic directional coupler~\cite{Sadovnikov2015,Wang2018}, proposed to be a key building block in wave-based logic and computing~\cite{Sadovnikov2016,Wang2020}. The magnon-cooparon based design that we demonstrate here offers stronger coupling strengths, smaller footprint, additional control (e.g., via temperature), and universality (e.g., for antiferromagnets~\cite{Kamra2018,Bobkov2021}) as compared to the dipole-interaction based designs considered previously~\cite{Sadovnikov2015,Wang2018}.

Considering the setup depicted in Fig.~\ref{Fig3}, we now assume existence of spin waves with the wavevector $k \mathbf{e}_z$ in both FIs, assumed identical for simplicity. As a result, there exists the same static exchange field $h_0$ in S below both the FI wires. However, distinct dynamic exchange fields $\delta h_{\mathrm{l,r}}$, similar to Eq.~\eqref{eq:h}, exist in S below each of the FI wires. These are proportional to the respective spin wave amplitudes $\delta m_{\mathrm{l,r}}$ in the two FIs. Solving the Usadel equation \eqref{eq:usadel} under this exchange field profile as detailed in the SM, we obtain the spin density in S: $\mathbf{s}(x) = \mathbf{s}_{\mathrm{l}}(x) + \mathbf{s}_{\mathrm{r}}(x)$ with
\begin{align}\label{eq:snl}
\mathbf{s}_{\mathrm{l,r}}(x) & = s_0(x) + s_{\mathrm{loc}}(x) \frac{\delta \mathbf{h}_{\mathrm{l,r}}}{h_0} + s_{\mathrm{nl}}(x) \frac{\delta \mathbf{h}_{\mathrm{r,l}}}{h_0}.
\end{align}
Here, the contributions $s_0$ and $s_{\mathrm{loc}}$ are due to the static and dynamic exchange fields induced by the FI directly above the S region. Thus, these are identical to our analysis of the magnon-cooparon in a FI/S bilayer. The nonlocal contribution $s_{\mathrm{nl}}(x)$ characterizes the spin density generated in S below the left FI by the right one, and vice versa. Relegating its detailed expression to the SM, we note that for $d \lesssim \xi_{\mathrm{S}}$, $s_{\mathrm{nl}}$ is comparable to the spin density $s_{\mathrm{loc}}$ accompanying a magnon-cooparon.

The induced nonlocal spin density leads to a fieldlike spin torque with the contribution $ \tilde{J} \bar{s}_{\mathrm{nl}} \mathbf{m}_{\mathrm{l,r}} \times \mathbf{m}_{\mathrm{r,l}}$ to the magnetization dynamics $\dot{\mathbf{m}}_{\mathrm{l,r}}$ in the two FIs~\cite{Ojajarviarxiv}, where $\bar{s}_{\mathrm{nl}}$ is $s_{\mathrm{nl}}(x)$ averaged over the width $t$ of the FI wire (Fig.~\ref{Fig3}), ad detailed further in the Supplementary Note 4. The resulting eigenmodes are magnon-cooparons distributed over the two FIs and the S layer with dispersion: $\omega_{\pm} = \gamma K + \tilde{D}_{\mathrm{m}} k^2 \mp \tilde{J} \bar{s}_{\mathrm{nl}}$. Hence, a pure spin wave injected with frequency $\omega$ into the left FI transfers its energy via the spinful superfluid to the right FI after traveling the so-called~\cite{Wang2020} coupling length $L$:
\begin{align}\label{eq:L}
L & = \frac{2 \pi}{k_+ - k_-} = \frac{\pi \tilde{D}_{\mathrm{m}} (k_+ + k_-)}{\tilde{J} \bar{s}_{\mathrm{nl}}},
\end{align}
where $k_{\pm}$ are the wavenumbers corresponding to the frequency $\omega$ of the injected spin wave. A smaller $L$ allows transfer of energy and the concomitant implementation of logic operations in smaller devices and thus is desirable.

\subsection{Numerical estimates and experimental detection}

We now employ material parameters pertinent to yttrium iron garnet~\cite{Xiao2010} as FI and Nb as S in finding the effects discussed above to be large. We consider $D_{\mathrm{m}} = 5 \cdot 10^{-29}~\mathrm{erg} \cdot \mathrm{cm}^2$, $M_{\mathrm{s}} = 140$ G, $\gamma = 1.76 \cdot 10^7~\mathrm{G}^{-1} \mathrm{s}^{-1}$, $\gamma K = 10^{-17}$ erg, $D = 3~\mathrm{cm}^2 \mathrm{s}^{-1}$, $N_{\mathrm{F}} = 1.3 \cdot 10^{35}~\mathrm{erg}^{-1} \mathrm{cm}^{-3}$, $\Delta_0 = 18$ K, and $\xi_{\mathrm{S}} = \sqrt{D/\Delta_0} \sim 10$ nm. Further, we consider $d_{\mathrm{FI}} = d_{\mathrm{S}}$. Figure \ref{Fig2} plotted with these values shows a large enhancement of the effective mass with decreasing temperature. This can be measured using, for example, the Brillouin Light Scattering technique~\cite{Demokritov2001} employed regularly in measuring magnon group velocities~\cite{Nembach2015}. Furthermore, the enhanced effective mass, and thus an altered spin conductivity, will manifest itself in the typical nonlocal magnonic spin transport experiment~\cite{Cornelissen2015,Goennenwein2015}. As the magnon spin conductivity scales as $\sim 1/\sqrt{D_{\mathrm{m}}}$~\cite{Cornelissen2016}, its fractional modification due to the magnon-cooparon formation is given by $- \delta D_{\mathrm{m}} / 2 D_{\mathrm{m}}$ and is expected to be large (Fig.~\ref{Fig2}).  Besides the in-situ control via, for example, temperature, the FI thickness can be used to engineer $\tilde{D}_{m}$ ex-situ. A negative value of $\tilde{D}_{\mathrm{m}}$ signifies that our assumed uniformly ordered magnetization is no longer the ground state~\cite{Bergeret2000}.

With the material parameters above, $h_0 = 0.61 \Delta_0$, and $T = 0.9 T_{\mathrm{c}}$, the net spin of the magnon-cooparon [Eq.~\eqref{eq:magcoopspin}] is evaluated as $0.4$, reduced from spin 1 of the bare magnon. Further, assuming $t = 10 \xi_{\mathrm{S}}$, $d = \xi_{\mathrm{S}}$, and $(k_+ + k_-)/2 = 10^7~\mathrm{m}^{-1}$, the coupling length $L$ [Eq.~\eqref{eq:L}] of the magnon-cooparon based directional coupler is evaluated as $\sim 100$ nm. This is an order of magnitude smaller than the coupling length afforded by dipolar-interaction based designs~\cite{Sadovnikov2015,Wang2018}. The experimental realization of the magnon-cooparon based directional coupler can follow the procedure similar to its dipole-interaction based counterpart~\cite{Sadovnikov2016,Wang2018} with the FI layers deposited on a superconductor instead of a substrate. Magnons with nonzero $k$ are often generated by applying ac voltage to a narrow conductor deposited on the FI~\cite{Sadovnikov2016,Wang2018}. The resulting spatially varying Oersted magnetic field bears a broadband $k$ spectrum and excites the finite-$k$ magnon that matches the frequency of the exciting voltage. Several other techniques can also generate finite-$k$ magnons by exploiting the same lack of $k$-conservation in hybrid systems~\cite{Kamra2015}.

\section{Conclusions}

We have demonstrated theoretically the ubiquitous existence of a quasiparticle, termed magnon-cooparon, comprising a spin flip in a magnetic insulator screened by a spinful Cooper pairs cloud in an adjacent conventional superconductor. The nonlocal nature of the magnon-cooparon is then exploited to propose a high performance magnonic directional coupler. While we have focused on a uniformly ordered ferromagnetic insulator, our analysis is general and anticipates an important role for magnon-cooparons in a wide range of hybrids comprising different magnetic insulators with various ground states.

\section{Methods}

Our theoretical method is a combination of Landau-Lifshitz-Gilbert (LLG) equation to describe the dynamics of the magnetization in the ferromagnetic part of the system and the nonequilibrium quasiclassical theory in terms of Usadel equations for Green's functions to describe the conductivity electrons in the superconducting part. The Green's function is used to calculate the electron spin polarization in the superconductor. The coupling between the LLG equation and the Usadel equation results from the S/FI interface exchange hamiltonian. It provides the spin torque term in the LLG equation, determined by the electron spin polarization in the superconductor.  Simultaneously the exchange hamiltonian gives rise to the exchange field term in the Usadel equation, which is generated by the FI magnetization.

The coupled system of the LLG and Usadel equations is solved analytically to obtain expressions for the magnon dispersion via the electron spin polarization and for the  quasiclassical Green's function via the magnetization profile. The self-consistency between these quantities is achieved numerically. The superconducting order parameter is also calculated self-consistently via the Green's function in the framework of this procedure. The closed analytical results for the renormalization  of a magnon stiffness and its spin were obtained in the limiting case of high temperatures $\Delta \ll T_{\mathrm{c}}$ and adiabatic approximation $\omega \ll T$.

\section{Data availability}

All data needed to evaluate the conclusions in the paper are present in the paper and/or
the Supplementary Material.


\section{Acknowledgments}

We acknowledge financial support from the Spanish Ministry for Science and Innovation -- AEI Grant CEX2018-000805-M (through the ``Maria de Maeztu'' Programme for Units of Excellence in R\&D), from the Deutsche Forschungsgemeinschaft (DFG; German Research Foundation) via a DFG-RSF project (ID 443404566), the SFB 1432 (ID 425217212), the SPP 2244 (ID 443404566) and from the Russian Science Foundation via the RSF-DFG project No.22-42-04408.

\section{Author contributions}

All authors contributed equally to the conception of the work, the analysis and interpretation of the results. I.V.B.~performed the analytical calculations. A.M.B. performed the numerical calculations. A.K. and I.V.B. wrote the manuscript with input from A.M.B and W.B.

\section{Competing interests}

The authors declare no competing interests.

\section{Supplemental information}

 \subsection{\!\!\!\!\!\!\!Supplementary Note 1: QUASICLASSICAL GREEN'S FUNCTION DESCRIPTION OF THE SUPERCONDUCTOR}

\label{formalism}

In the superconductor the Usadel equation for the $8 \times 8$ matrix Green's function $\check g$ in the direct product of Keldysh, spin and particle-hole spaces takes the form:
\begin{eqnarray}
\mathrm i D \mathbf{\nabla} \bigl( \check g \otimes \mathbf{\nabla} \check g \bigr) = \bigl[ \varepsilon \hat \tau_z - \mathbf{h} \hat {\bm \sigma} \hat \tau_z + \Delta \mathrm i \hat \tau_y, \check g \bigr]_\otimes,
\label{usadel_1}
\end{eqnarray}
where $[A,B]_\otimes = A\otimes B -B \otimes A$ and we work in the mixed $(\varepsilon,t)$ representation with $A \otimes B = \exp[(i/2)(\partial_{\varepsilon_1} \partial_{t_2} -\partial_{\varepsilon_2} \partial_{t_1} )]A(\varepsilon_1,t_1)B(\varepsilon_2,t_2)|_{\varepsilon_1=\varepsilon_2=\varepsilon;t_1=t_2=t}$. In case $ A[B](\varepsilon, t) =  A[B] (\varepsilon) \exp[i \Omega_{A[B]} t]$ the $\otimes$-product is reduced to $ A(\varepsilon,t) \otimes B(\varepsilon, t) = A (\varepsilon-\Omega_B/2)  B(\varepsilon+\Omega_A/2,t)\exp[i (\Omega_A + \Omega_B) t]$. $\hat \tau_{x,y,z}$ and $\hat \sigma_{x,y,z}$ are Pauli matrices in particle-hole and spin spaces, respectively. $\Delta $ is the superconducting order parameter. The explicit structure of the Green's function in the Keldysh space takes the form:
\begin{eqnarray}
	\check g =
	\left(
	\begin{array}{cc}
	\breve g^{\mathrm R} & \breve g^{\mathrm K} \\
	0 & \breve g^{\mathrm A}
	\end{array}
	\right),
	\label{keldysh_structure}
	\end{eqnarray}
	where $\breve g^{{\mathrm R}({\mathrm A})}$ is the retarded (advanced) component of the Green's function and $\breve g^{\mathrm K}$ is the Keldysh component. Further we express the Keldysh part of the Green's function  via the retarded, advanced Green's function and the distribution function $\breve \varphi$ as follows: $\breve g^{\mathrm K} = \breve g^{\mathrm R} \otimes \breve \varphi - \breve \varphi \otimes \breve g^{\mathrm A}$.

The exchange field is taken in the form of a time-independent component and a circularly polarized magnon:
\begin{eqnarray}
\mathbf{h} = h_0 \mathbf{e}_z + \delta h \cos(\mathbf{k} \mathbf{r} + \omega t) \mathbf{e}_x + \delta h \sin(\mathbf{k} \mathbf{r} + \omega t) \mathbf{e}_y .
\label{h}
\end{eqnarray}
Then
\begin{eqnarray}
\delta \mathbf{h} \bm \sigma = \delta h {\mathrm e}^{-\mathrm i(\mathbf{k} \mathbf{r} + \omega t)\hat \sigma_z}\hat \sigma_x .
\label{h2}
\end{eqnarray}
The quasiclassical Green's function is to be found in the form: $\check g = \check g_0 + \delta \check g$, where $\check g_0$ is the Green's function in the absence of the magnon and $\check \delta g$ is the first order correction with respect to $\delta \mathbf{h}$. Taking into account that $\mathbf{\nabla} \check g_0 = 0$ (we assume that in the absence of the magnon the bilayer is spatially homogeneous along the interface) from Eq.~(\ref{usadel_1}) we obtain the following equation for $\delta \check g$:
\begin{eqnarray}
\mathrm i D \check g_0 \otimes \nabla^2  \delta \check g = \bigl[  \varepsilon \hat \tau_z - h_0 \mathbf{e}_z \hat \sigma_z \hat \tau_z + \hat \tau_z \hat \Delta, \delta \check g \bigr]_\otimes - \bigl[  \delta h {\mathrm e}^{-\mathrm i(\mathbf{k} \mathbf{r} + \omega t)\hat \sigma_z} \hat \sigma_x \hat \tau_z, \check g_0 \bigr]_\otimes
\label{delta_g1}
\end{eqnarray}
Introducing the unitary operator $\hat U =  {\mathrm e}^{-\mathrm i(\mathbf{k} \mathbf{r} + \omega t)\hat \sigma_z/2}$ we can transform the  Green's function as follows:
\begin{eqnarray}
\delta \check g = \hat U \otimes \delta \check g_{\mathrm m} \otimes \hat U^{\dagger} .
\label{transform}
\end{eqnarray}
In case if the system is spatially homogeneous except for the magnon, $\delta \check g_{\mathrm m}$ does not depend on coordinates. Then
\begin{eqnarray}
\nabla^2  \delta \check g = -\frac{k^2}{2} \hat U \otimes \bigl( \delta \check g_{\mathrm m} - \hat \sigma_z \delta \check g_{\mathrm m} \hat \sigma_z  \bigr) \otimes \hat U^\dagger = -k^2 \delta \check g,~~~~
\label{delta_g2}
\end{eqnarray}
where when passing to the second equality it is used that $\delta \check g = \delta \breve g_x \hat \sigma_x + \delta \breve g_y \hat \sigma_y$ and has no $z$-component in the spin space according to the spin structure of the magnon exchange field $\delta \mathbf{h}$.

From the normalization condition $\check g \otimes \check g = 1$ it follows that $\check g_0 \otimes \delta \check g = - \delta \check g \otimes \check g_0$. It gives us $\check g_0 \otimes \delta \check g = (1/2)[\check g_0, \delta \check g]_\otimes$. Eq.~(\ref{delta_g1}) takes the form:
\begin{eqnarray}
\bigl[  \varepsilon \hat \tau_z  + \mathrm i \frac{D k^2}{2} \check g_0 - h_0  \hat \sigma_z \hat \tau_z  + \hat \tau_z \hat \Delta, \delta \check g \bigr]_\otimes - \bigl[  \delta h {\mathrm e}^{-\mathrm i(\mathbf{k} \mathbf{r} + \omega t)\hat \sigma_z} \hat \sigma_x \hat \tau_z , \check g_0 \bigr]_\otimes = 0.
\label{delta_g3}
\end{eqnarray}
From Eq.~(\ref{delta_g3}) the following equation for  $\delta \check g_m$ is obtained:
\begin{eqnarray}
\bigl[ \breve \Lambda_{\mathrm{d}} \hat \tau_z + \breve \Lambda_{\mathrm{od}} \mathrm i\hat \tau_y, \delta \check g_{\mathrm m} \bigr] = \delta h \bigl[ \hat \sigma_x \hat \tau_z, \nonumber \\
\frac{1}{2}\bigl((\hat g_{0,+}+\hat g_{0,-}\hat \sigma_z)\hat \tau_z + (\hat f_{0,+}+\hat f_{0,-}\hat \sigma_z)\mathrm i\hat \tau_y \bigr) \bigr].
\label{delta_g4}
\end{eqnarray}
It does not contain time dependence and $\otimes$-products. In Eq.~(\ref{delta_g4}) we use the following definitions
\begin{eqnarray}
\hat g_{0,\pm} = \hat g_{0,\uparrow}(\varepsilon+\frac{\omega}{2}) \pm \hat g_{0,\downarrow}(\varepsilon-\frac{\omega}{2}), \label{delta_g_def1} \\
\hat f_{0,\pm} = \hat f_{0,\uparrow}(\varepsilon+\frac{\omega}{2}) \pm \hat f_{0,\downarrow}(\varepsilon-\frac{\omega}{2}),
\label{delta_g_def2}
\end{eqnarray}
where $\hat g_{0,\uparrow(\downarrow)}$ represent the bulk Green's functions for the superconductor in the exchange field $h_0$:
\begin{eqnarray}
g_{0,\uparrow(\downarrow)}^{\mathrm R} = \frac{|\varepsilon \mp h_0|}{\sqrt{(\varepsilon + \mathrm i \delta \mp h_0)^2-\Delta^2}} \label{delta_g_def3} \\
f_{0,\uparrow(\downarrow)}^{\mathrm R} = \frac{\Delta {\rm sgn}(\varepsilon \mp h_0)}{\sqrt{(\varepsilon + \mathrm i \delta \mp h_0)^2-\Delta^2}},
\label{delta_g_def4}
\end{eqnarray}
and $g(f)_{0,\uparrow(\downarrow)}^{\mathrm A} = -g(f)_{0,\uparrow(\downarrow)}^{{\mathrm R}*}$, $g(f)_{0,\uparrow(\downarrow)}^{\mathrm K} = [g(f)_{0,\uparrow(\downarrow)}^{\mathrm R} - g(f)_{0,\uparrow(\downarrow)}^{\mathrm A}] \tanh[\varepsilon/2T]$.
\begin{eqnarray}
\breve \Lambda_{\mathrm{d}} = \hat \Lambda_{\mathrm{d}}^0 + \hat \Lambda_{\mathrm{d}}^z \hat \sigma_z = \varepsilon + \frac{\mathrm i D k^2}{4}\hat g_{0,+} + \bigl( \frac{\omega}{2}-h_0 + \frac{\mathrm i D k^2}{4}\hat g_{0,-} \bigr)\hat \sigma_z,
\label{delta_g_def5} \\
\breve \Lambda_{\mathrm{od}} = \hat \Lambda_{\mathrm{od}}^0 + \hat \Lambda_{\mathrm{od}}^z \hat \sigma_z = \Delta + \frac{\mathrm i D k^2}{4}\hat f_{0,+} + \frac{\mathrm i D k^2}{4}\hat f_{0,-} \hat \sigma_z .
\label{delta_g_def6}
\end{eqnarray}
Solving Eq.~(\ref{delta_g4}) we obtain:
\begin{eqnarray}
\delta \check g_{\mathrm m} = \delta \hat g_{\mathrm mx} \hat \sigma_x \hat \tau_z + \delta \hat f_{\mathrm mx} \hat \sigma_x \mathrm i\hat \tau_y ,
\label{delta_g_sol1}
\end{eqnarray}
where
\begin{eqnarray}
\delta \hat g_{\mathrm mx} = \frac{\delta h \bigl[ \hat f_{0,+}\hat \Lambda_{\mathrm{od}}^z - \hat g_{0,-}\hat \Lambda_{\mathrm d}^0 \bigr]}{2 \bigl[ \hat \Lambda_{\mathrm d}^0 \hat \Lambda_{\mathrm d}^z - \hat \Lambda_{\mathrm {od}}^0\hat \Lambda_{\mathrm {od}}^0 \bigr]}
\label{delta_g_sol2}
\end{eqnarray}
The distribution function also acquires a correction due to the magnon: $\breve \varphi = \breve \varphi_0 + \delta \check \varphi$. It is convenient to work with the transformed distribution function $\hat U^\dagger \breve \varphi \hat U = \breve \varphi_{\mathrm m} + \delta \breve \varphi_{\mathrm m}$, which does not depend on time and spatial coordinates. Here $\breve \varphi_{\mathrm m} = (1/2)[\varphi_{\mathrm m+}+\varphi_{\mathrm m-}\sigma_z]\hat \tau_0$ with $\varphi_{\mathrm m\pm} = \tanh[(\varepsilon+\omega/2)/2T] \pm \tanh[(\varepsilon-\omega/2)/2T]$ is the result of the unitary transformation of the equilibrium distribution function $ \varphi_0 = \tanh [\varepsilon/2T]$. From the Keldysh part of the Usadel equation (\ref{delta_g4}) we can obtain the following equation for the first order correction to the distribution function $\delta \check \varphi_{\mathrm m} $:
\begin{eqnarray}
\mathrm i Dk^2 [\delta \breve \varphi_{\mathrm m} - \breve g_{{\mathrm m}0}^{\mathrm R} \delta \breve \varphi_{\mathrm m} \breve g_{{\mathrm m}0}^{\mathrm A}] + \breve g_{{\mathrm m}0}^{\mathrm R} [\breve K, \delta \breve \varphi_{\mathrm m}] - [\breve K, \delta \breve \varphi_{\mathrm m}] \breve g_{{\mathrm m}0}^{\mathrm A} + \nonumber \\
	\breve g_{{\mathrm m}0}^{\mathrm R} [\breve \varphi_{\mathrm m}, \delta h \hat \sigma_x \hat \tau_z] - [\breve \varphi_{\mathrm m}, \delta h \hat \sigma_x \hat \tau_z] \breve g_{{\mathrm m}0}^{\mathrm A} = 0,
\label{distribution_correction}
\end{eqnarray}
where
\begin{eqnarray}
\breve K = (\varepsilon+(\omega/2-h_0)\hat \sigma_z )\hat \tau_z +  \Delta \mathrm i \hat \tau_y,
\label{distribution_def_1}
\end{eqnarray}
\begin{eqnarray}
\breve g_{{\mathrm m}0}^{\mathrm R,\mathrm A} = \hat U^\dagger \otimes \breve g_0^{\mathrm R,\mathrm A} \otimes \hat U = (1/2)[(g_{0,+}^{\mathrm R,\mathrm A}+g_{0,-}^{\mathrm R,\mathrm A}
\hat \sigma_z)\hat \tau_z + (f_{0,+}^{\mathrm R,\mathrm A}+f_{0,-}^{\mathrm R,\mathrm A}\hat \sigma_z)\mathrm i\hat \tau_y],
\label{distribution_def_2}
\end{eqnarray}
Structure of Eq.~(\ref{distribution_correction}) dictates that
\begin{eqnarray}
\delta \check \varphi_{\mathrm m} =
\left(
\begin{array}{cc}
0 & \delta \varphi_{\mathrm m}^\uparrow \\
\delta \varphi_{\mathrm m}^\downarrow & 0
\end{array}
\right)\hat \tau_0.
\label{distribution_correction_2}
\end{eqnarray}
Substituting Eq.~(\ref{distribution_correction_2}) into Eq.~(\ref{distribution_correction}) we obtain the following result:
\begin{eqnarray}
\delta \varphi_{\mathrm m}^\sigma = -\delta h \varphi_{{\mathrm {m-}}} \frac{2h_{0\omega}G_{-,\sigma}+\mathrm i \sigma D k^2 (g_{{\mathrm m}\sigma}^{\mathrm R} - g_{{\mathrm m} \bar \sigma}^{\mathrm A})}{4h_{0\omega}^2 G_{-,\sigma} + 4 h_{0\omega} {\mathrm i} \sigma D k^2 (g_{{\mathrm m}\sigma}^{\mathrm R} - g_{{\mathrm m} \bar \sigma}^{\mathrm A}) - (Dk^2)^2 G_{+,\sigma}},~~~~
\label{distribution_correction_3}
\end{eqnarray}
where we introduce the spin subband index $\sigma = \uparrow(\downarrow)$ in the subscripts/superscripts and $\sigma = \pm 1$ for spin-up(down) subbands, respectively, if it is as a factor. $\bar \sigma = -\sigma$, $h_{0\omega} = \omega/2-h_0$, ${g(f)}_{{\mathrm m}\sigma}^{{\mathrm R},{\mathrm A}} = g(f)_{0,\sigma}^{{\mathrm R},{\mathrm A}}(\varepsilon + \sigma \omega/2)$ and $G_{\pm,\sigma} = 1-g_{{\mathrm m}\sigma}^{\mathrm R} g_{{\mathrm m} \bar \sigma}^{\mathrm A} \pm f_{{\mathrm m}\sigma}^{\mathrm R} f_{{\mathrm m} \bar \sigma}^{\mathrm A}$.

\newpage

\subsection{\!\!\!\!\!\!\!Supplementary Note 2: RENORMALIZATION OF THE EXCITATION DISPERSION AND DAMPING}

\label{sec:dispersion}

Following the methodology detailed in the previous section, the electron spin polarization $\bm s$ in the superconductor can be calculated as
\begin{eqnarray}
\mathbf{s} = -\frac{N_{\mathrm F}}{16} \int d\varepsilon {\rm Tr}_4 \Bigl[ \hat {\bm \sigma} \hat \tau_z \breve g^{\mathrm K} \Bigr].
\label{s}
\end{eqnarray}
It can be written in the form:
\begin{eqnarray}
\mathbf{s} = s_0 \mathbf{m}_0 + \delta s_\parallel \delta \mathbf{m} + \delta s_\perp (\delta \mathbf{m} \times \mathbf{m}_0),
\label{spin2}
\end{eqnarray}
where $s_0$ is the equilibrium value of the electron spin polarization in the superconductor, corresponding to the absence of the magnon. $\delta s_\parallel$ and $\delta s_\perp$ describe the dynamic corrections to the spin polarization due to the magnon.
\begin{eqnarray}
s_0  = -\frac{N_{\mathrm F}}{4} \int \limits_{-\infty}^\infty d \varepsilon \tanh \frac{\varepsilon}{2T} {\rm Re}\bigl[ g_{0,\uparrow}^{\mathrm R} - g_{0,\downarrow}^{\mathrm R} \bigr],
\label{spin_0}
\end{eqnarray}
\begin{eqnarray}
\delta s_\parallel  = -\frac{N_{\mathrm F}h_0}{8\delta h} \int \limits_{-\infty}^\infty d \varepsilon \Bigl\{ 2 \varphi_{\mathrm m+} {\rm Re} [\delta g_{\mathrm mx}^{\mathrm R}] +  \sum \limits_\sigma (g_{\mathrm m\sigma}^{\mathrm R} - g_{\mathrm m \bar \sigma}^{\mathrm A}) \delta \varphi_{\mathrm m}^\sigma  \Bigr\},
\label{spin_parallel}
\end{eqnarray}
\begin{eqnarray}
\delta s_\perp  = \frac{N_{\mathrm F}h_0}{8\delta h} \int \limits_{-\infty}^\infty d \varepsilon \Bigl\{ 2 \varphi_{{\mathrm m}-} {\rm Im} [\delta g_{{\mathrm m}x}^{\mathrm R}] +{\mathrm i}  \sum \limits_\sigma \sigma (g_{{\mathrm m}\sigma}^{\mathrm R} - g_{{\mathrm m} \bar \sigma}^{\mathrm A}) \delta \varphi_{\mathrm m}^\sigma\Bigr\}.
\label{spin_perpendicular}
\end{eqnarray}
As per the main text, $\delta s_\parallel - s_0$ accounts for the renormalization of the magnon dispersion. It can be shown that in the adiabatic limit $\hbar \omega \ll T$ if one neglects $\hbar \omega $ with respect to the superconducting energies $\delta s_\parallel - s_0 \propto Dk^2$. Consequently, it only renormalizes the magnon stiffness. The explicit expression for the stiffness correction in the limit $T \to T_{\mathrm c}$ takes the form
\begin{eqnarray}
\delta D_{\mathrm m} = -\frac{ \pi T_{\mathrm c} N_{\mathrm F} \hbar \gamma d_{\mathrm S} D}{d_{\mathrm {FI}} M_{\mathrm s}} \sum \limits_{\omega_n > 0} \frac{\Delta^2 h_0^2}{[\omega_n^2 + h_0^2]^2 }.
\label{stiffness_renorm_lin}
\end{eqnarray}
The same result for the stiffness renormalization has been obtained from the consideration of the total energy of the bilayer in the framework of the Ginzburg-Landau theory~\cite{Silaevarxiv}.

If one does not make the adiabatic approximation, then the zero-momentum value of the magnon energy is also slightly renormalized. The renormalization correction takes the form:
\begin{eqnarray}
\hbar (\omega_{\mathbf{k} = 0} - \gamma K) =  \frac{J}{d_{\mathrm {FI}}} \bigl(\delta s_\parallel(\omega,\mathbf{k}=0)-\delta s_\parallel(\omega=0,\mathbf{k}=0)\bigr),
\label{renorm_zero_momentum}
\end{eqnarray}
where
\begin{eqnarray}
\delta s_\parallel(\omega,\mathbf{k}=0) = \frac{N_{\mathrm F}}{4} \frac{h_0}{\omega/2-h_0} \int \limits_{-\infty}^\infty d \varepsilon [\tanh \frac{\varepsilon+\frac{\omega}{2}}{2T}{\rm Re}[g_{0,\uparrow}^{\mathrm R}(\varepsilon+\frac{\omega}{2})] - \nonumber \\
\tanh \frac{\varepsilon-\frac{\omega}{2}}{2T}{\rm Re}[g_{0,\downarrow}^{\mathrm R}(\varepsilon-\frac{\omega}{2})]].
\label{renorm_zero_momentum_2}
\end{eqnarray}
The renormalization of the zero-momentum magnon frequency in the limit of $h_0 \to 0$ has already discussed~\cite{Silaev2020}, and our result above is consistent with this previous work in the relevant limit $h_0 \to 0$. The general results obtained above have been employed in plotting the solid lines of Fig.~2 in the main text.

For estimates and numerical calculations we take material parameters of YIG as a FI and Nb as a superconductor. $\sigma_{\mathrm S} = 9 \cdot 10^{16}s^{-1}$ is the Nb conductivity and $D/\hbar = 3 cm^2 \cdot s^{-1}$ is the Nb diffusion constant. Then the DOS at the Fermi level in Nb is $N_{\mathrm F} = \hbar \sigma_S/(e^2 D) = 0.13 \cdot 10^{36} erg^{-1}\cdot cm^{-3}$. The YIG parameters are \cite{Xiao2010} $D_{\mathrm m} = 5 \cdot 10^{-29}erg \cdot cm^2$, $M_{\mathrm s} = 1.4 \cdot 10^2 G$, $\gamma = 1.76 \cdot 10^7 G^{-1}s^{-1}$ and $\hbar \gamma K = 10^{-17} erg = 4 \cdot 10^{-3} \Delta_0$, where $\Delta_0 = 18 K$ is the zero-temperature order parameter in Nb. With these parameters we obtain that the renormalization of the zero-momentum magnon frequency $(\omega_{\mathbf{k} = 0} - \gamma K)/\gamma K \lesssim 10^{-2}$ due to the smallness of the bare magnon frequency $\gamma K$ with respect to $\Delta_0$.

\begin{figure}[tb]
	\begin{center}
		\includegraphics[width=85mm]{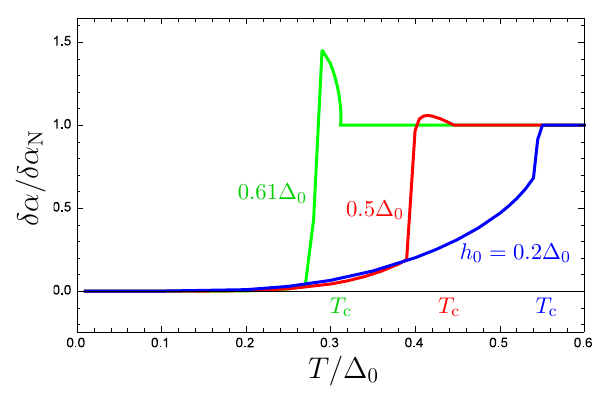}
		\caption{\textbf{Fractional increase in the efffective Gilbert damping for the magnon-cooparon due to the superconducting layer.} This enhancement is mediated by the quasiparticles, and not the superfluid condensate.} \label{fig:damping}
	\end{center}
\end{figure}

According to Eq.~(3) in the main text, $\delta s_\perp$ determines the correction to the magnon decay rate. It is zero in the framework of the quasistatic approximation when we neglect $\hbar \omega$ with respect to the superconducting energies. Beyond this approximation is it non-zero and the correction to the Gilbert damping parameter $\delta \alpha = -(J/d_{\mathrm {FI}})\delta s_\perp/\omega$ is presented in Fig.~\ref{fig:damping} as a function of temperature for different values of $h_0$. The normal state value of $\delta \alpha$ can be obtained analytically:
\begin{eqnarray}
\delta \alpha_{\mathrm N} =  \frac{d_{\mathrm S}}{d_{\mathrm {FI}}}\frac{(2\hbar \gamma N_{\mathrm F} h_0^2/M_{\mathrm s}) Dk^2}{4(\frac{\omega}{2}-h_0)^2 + (Dk^2)^2} .
\label{gilbert_correction_normal}
\end{eqnarray}
It is seen that $\delta \alpha_{\mathrm N}$ is positive and vanishes for zero-momentum magnons. The result is natural because in the framework of our model we consider the only source of spin relaxation processes in superconductor - the finite momentum of the magnon, which results in the spin relaxation rate $Dk^2$.

\begin{figure}[tb]
	\begin{center}
		\includegraphics[width=85mm]{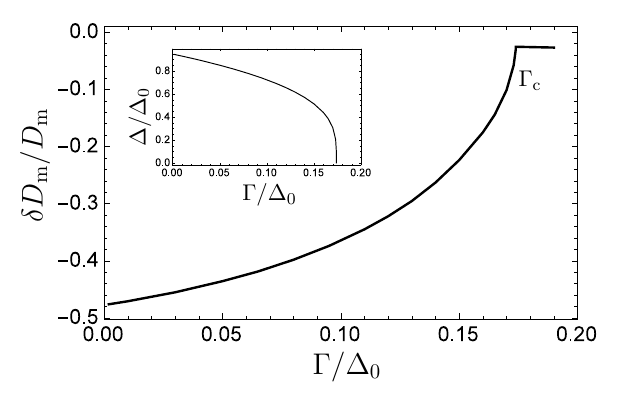}
		\caption{\textbf{Dependence of the stiffness correction on the Dynes parameter.} $h_0=0.5\Delta_0$, $T=0.2\Delta_0$. Insert: Superconducting order parameter as a function of the Dynes parameter. $\Gamma_{\mathrm c}$ is the critical value of the Dynes parameter, which fully suppresses superconductivity for a given $h_0$ and $T$.} \label{fig:Gamma}
	\end{center}
\end{figure}

We do not consider any other spin relaxation processes such as spin-orbit relaxation and relaxation at magnetic impurities. The corresponding relaxation rates would additionally increase the correction to the Gilbert damping. Also the spin-flip scattering suppresses superconducting order parameter \cite{Abrikosov1962}. In its turn, the suppression of the order parameter leads to the suppression of the triplets, which are generated from the singlets. This fact results in weakening of the magnon screening by the triplet cloud. To support these qualitative arguments, in Fig.~\ref{fig:Gamma} we have plotted the dependence of the renormalized stiffness on the Dynes parameter $\Gamma$, which roughly models the effect of  spin-flip scattering on the spectral functions and order parameter. We can also roughly estimate the realistic value of $\Gamma \sim \tau_{\mathrm s}^{-1} $, where $\tau_{\mathrm s}$ is the spin flip scattering time. Taking $\tau_{\mathrm s} \sim 25-100ps$, as it was reported for Al thin films \cite{Poli2008} we obtain $\Gamma \sim 10^{-1}K \sim 10^{-1}\Delta_{0,{\mathrm {Al}}}$ or $\Gamma \sim 10^{-2}\Delta_{0,\mathrm {Nb}}$, where $\Delta_{0,\mathrm {Al(Nb)}}$ is the superconducting order parameter at zero temperature for Al(Nb) superconductor. In our calculations we focus on Nb parameters.

The correction to the Gilbert damping due to the spin-orbit relaxation and relaxation at magnetic impurities has been investigated in Ref.~\onlinecite{Silaev2020} by calculation of the spin susceptibility of the superconductor. Eq.~(\ref{gilbert_correction_normal}) at $h_0 \to 0$ coincides with the result obtained in Ref.~\onlinecite{Silaev2020} with the substitution $Dk^2 \to 2\Gamma + \tau_{\mathrm s}^{-1}$, where $\Gamma$ is the Dynes parameter and $\tau_{\mathrm s}^{-1}$ is the spin-orbit relaxation rate. Our estimates suggest that $\delta \alpha_{\mathrm N} \sim 10^{-2} (\xi_{\mathrm S} k^2)$, where $\xi_{\mathrm S} = \sqrt{D/\Delta_0} $ is the superconducting coherence length, which is of the order of $10$ nm for Nb films. Consequently, for short wave-length magnons the superconducting correction is essential in comparison with the extremely low values of the Gilbert damping parameter in YIG\cite{Xiao2010} $\alpha \approx 5 \cdot 10^{-5}$. The influence of superconductivity on the correction to the Gilbert damping constant is presented in Fig.~\ref{fig:damping}. The most important feature of the temperature dependence is that in the presence of the finite $h_0$ in the superconductor the correction in the superconducting state demonstrates a sharp drop at a given temperature, which physically corresponds to the value of the superconducting order parameter $\Delta (T) = h_0$, where the gap in the superconducting state is closed. At this point the number of thermal quasiparticles, which can flip their spin, grows sharply. The results obtained at $h_0 \to 0$ and, consequently, not taking into account the Zeeman splitting of the superconducting gap, manifest gradual decline of the correction at the superconducting state \cite{Silaev2020}.

\newpage
\subsection{\!\!\!\!\!\!\!Supplementary Note 3: SPIN CURRENT}

Now we calculate the spin current carried by the triplet pairs, induced in the superconductor by the magnon:
\begin{eqnarray}
\hat j_{\mathrm s} = \frac{\hbar N_{\mathrm F} D}{16}{\rm Tr}_4 \int d \varepsilon \hat {\bm \sigma} \bigl( \check g {\mathbf {\nabla}} \check g \bigr)^{\mathrm K}
\label{spin_current_1}
\end{eqnarray}
The first order contribution with respect to $\delta h$ to the spin current is zero after the time averaging. Therefore, the spin current is of the second order with respect to $\delta h$. It has the only non-zero component $j_{S,z}$ carrying the $z$-component of spin along the $\bm k$ direction. In general, the calculation of second order terms with respect to $\delta h$ terms in the Green's function is rather cumbersome. Here we restrict ourselves by the calculation at $T \to T_{\mathrm c}$, where the Green's functions can be linearized with respect to the superconducting order parameter $\Delta$. Then the solution for the retarded quasiclassical Green's function takes the form $\breve g^{\mathrm R} = \hat \tau_z \hat \sigma_0 + \hat f^{\mathrm R} \hat \tau_+ + \hat {\tilde f}^{\mathrm R} \hat \tau_-$. The linearized with respect to the anomalous Green's function version of Eq.~(\ref{spin_current_1}) takes the form:
\begin{align}
\hat j_{\mathrm s} = & \frac{\hbar N_{\mathrm F} D}{16}{\rm Tr}_2 \int d \varepsilon \bm \sigma \bigl[ \frac{1}{2}\bigl(\hat f^{\mathrm R} \mathbf {\nabla} \hat {\tilde f}^{\mathrm R} + \hat {\tilde f}^{\mathrm R} \mathbf {\nabla} \hat f^{\mathrm R} - (\mathbf {\nabla} \hat f^{\mathrm R})  \hat {\tilde f}^{\mathrm R} - (\mathbf {\nabla} \hat {\tilde f}^{\mathrm R}) \hat f^{\mathrm R} - {\mathrm R} \to {\mathrm A}\bigr)\tanh \frac{\varepsilon}{2T} \nonumber \\
   & + 2 \mathbf {\nabla} \hat \varphi^{(2)}  + 2 \mathbf {\nabla} \hat {\tilde \varphi}^{(2)}\bigr],
\label{spin_current_2}
\end{align}
where the distribution function $\breve \varphi = \tanh[\varepsilon/2T]+\hat \varphi^{(2)}(1+\tau_z)/2 + \hat {\tilde \varphi}^{(2)}(1-\tau_z)/2$ contains the seconds order correction with respect to the anomalous Green's function due to the non-equilibrium generation of the triplet pairs.  At first let us calculate $\hat \varphi^{(2)}$. The equation for the distribution function is obtained from the Keldysh part of Eq.~(\ref{usadel_1}) and up to the second order with respect to the anomalous Green's function takes the form:
\begin{eqnarray}
\mathrm i D \mathbf {\nabla} \Bigl[ \frac{1}{2}\bigl( \hat f^{\mathrm R} \mathbf {\nabla} \hat {\tilde f}^{\mathrm R} -(\mathbf {\nabla} \hat f^{\mathrm R}) \hat {\tilde f}^{\mathrm R} - {\mathrm R} \to {\mathrm A} \bigr)\tanh \frac{\varepsilon}{2T} +  \nonumber \\
2 \nabla \hat \varphi^{(2)} \Bigr] = \frac{h_0}{2}[\hat \sigma_z, (\hat f^{\mathrm R} \hat {\tilde f}^{\mathrm R} + \hat f^{\mathrm A} \hat {\tilde f}^{\mathrm A})]\tanh \frac{\varepsilon}{2T} + ~~~~~~ \nonumber \\
\frac{1}{2}[\delta \mathbf{h} \hat {\bm \sigma}, (\hat f^{\mathrm R} \hat {\tilde f}^{\mathrm R} +
\hat f^{\mathrm A} \hat {\tilde f}^{\mathrm A})]\tanh \frac{\varepsilon}{2T} -2h_0 [\hat \sigma_z, \hat \varphi^{(2)}] - 2[\delta \mathbf{h} \hat {\bm \sigma},\hat \varphi^{(2)}].
\label{distribution_1}
\end{eqnarray}
In order to simplify the calculations further we work in the quasistatic approximation, when we neglect the correction of the order of $\omega/T_{\mathrm c}$ and higher orders of this parameter. In this limit all the $\otimes$-products in Eq.~(\ref{distribution_1}) can be changed  by the usual multiplication because the neglected corrections are of the order of $\omega/T_{\mathrm c}$, which is assumed to be small. To this approximation the correction $ \varphi_z^{(2)}$ to the distribution function (which could contribute to the spin current) is zero.

Then the spin current is expressed by the first two lines of Eq.~(\ref{spin_current_2}). In order to calculate the spin current we need to calculate
\begin{eqnarray}
\hat {\mathbf{I}} = \hat f^{\mathrm R} \mathbf {\nabla} \hat {\tilde f}^{\mathrm R} + \hat {\tilde f}^{\mathrm R} \mathbf {\nabla} \hat f^{\mathrm R} - (\mathbf {\nabla} \hat f^{\mathrm R})  \hat {\tilde f}^{\mathrm R} - (\mathbf {\nabla} \hat {\tilde f}^{\mathrm R}) \hat f^{\mathrm R} = -2[\hat f^{\mathrm R} \mathbf {\nabla} \hat {f}^{\mathrm R} - (\mathbf {\nabla} \hat f^{\mathrm R})  \hat { f}^{\mathrm R} ]
\label{spin_current_3}
\end{eqnarray}
the second order with respect to $\delta h$ contribution to $\hat {\mathbf{I}}$ takes the form:
\begin{eqnarray}
\hat {\mathbf{I}}^{(2)} = \mathrm i \mathbf {k} \hat U\Bigl[ \delta \hat f_{\mathrm m}^{\mathrm R} [\hat \sigma_z, \delta \hat f_{\mathrm m}^{\mathrm R} ] - [\hat \sigma_z, \delta \hat f_{\mathrm m}^{\mathrm R} ] \delta \hat f_{\mathrm m}^{\mathrm R} \Bigr]\hat U^\dagger
\label{spin_current_4}
\end{eqnarray}
The anomalous Green's function is $\hat f^{\mathrm R} = \hat f_0^{\mathrm R} + \delta \hat f^{\mathrm R}$, where $\delta \hat f^{\mathrm R} = \hat U \delta \hat f_{\mathrm m}^{\mathrm R} \hat U^\dagger$. $\delta \hat f_{\mathrm m}^{\mathrm R}$ can be obtained from Eq.~(\ref{delta_g4}) and takes the form:
\begin{eqnarray}
\delta \hat f_{\mathrm m}^{\mathrm R} = \delta f_{{\mathrm m},x}^{\mathrm R} \hat \sigma_x, \nonumber \\
\delta f_{{\mathrm m},x}^{\mathrm R} = \frac{\delta h f_{0,{\mathrm s}}^{\mathrm R}}{\varepsilon + \mathrm i D k^2/2}.
\label{anomalous_1}
\end{eqnarray}
In Eq.~(\ref{anomalous_1}) $f_{0,{\mathrm s}}^{\mathrm R} = \Delta \varepsilon/(\varepsilon^2-h_0^2)$ and $\varepsilon$ has an infinitesimal positive imaginary part $\delta$.

Substituting Eq.~(\ref{spin_current_4}) into Eq.~(\ref{spin_current_2}) and taking into account that $\hat f^{{\mathrm A}} = -\hat f^{{\mathrm R}*}$, we obtain:
\begin{eqnarray}
\mathbf{j}_{\mathrm S,z} = \frac{\hbar N_{\mathrm F} D\mathbf{k}}{2}  \int d \varepsilon \tanh \frac{\varepsilon}{2T} {\rm Im}\Bigl[ {\delta f_{{\mathrm m},x}^{\mathrm R}}^2 \Bigr]=
\nonumber \\ 2 \pi T_{\mathrm c} \hbar N_{\mathrm F} D\mathbf{k} \sum \limits_{\omega_n>0} \frac{\Delta^2 \delta h^2 \omega_n^2}{(\omega_n^2 + h_0^2)^2(\omega_n + Dk^2/2)^2}.
\label{spin_current_5}
\end{eqnarray}
The spin current in the FI, carried by the magnons takes the form:
\begin{eqnarray}
\mathbf{j}_{\mathrm m} = -\mathbf{k}  \Bigl( \frac{\delta h}{h_0} \Bigr)^2 \frac{\tilde D_{\mathrm m} M_{\mathrm s}}{\hbar \gamma},
\label{estimates_1}
\end{eqnarray}
where $\tilde D_{\mathrm m} = D_{\mathrm m} + \delta D_{\mathrm m}$ is the exchange stiffness renormalized by the superconductor and $M_{\mathrm s}$ is the saturation magnetization. Taking the YIG parameters \cite{Xiao2010} $D_{\mathrm m} = 5 \cdot 10^{-29}erg \cdot cm^2$, $\gamma = 1.76 \cdot 10^7 G^{-1} \cdot s^{-1}$ and $M_{\mathrm s} = 1.4 \cdot 10^2 G$, we obtain
\begin{eqnarray}
\bm j_{\mathrm m} \approx \left( \frac{\delta h}{h_0} \right)^2 \mathbf{k} \cdot 4 \cdot 10^{-7} erg \cdot cm^{-1}.
\label{estimates_2}
\end{eqnarray}
From the spin currents thus obtained, the net spin of the magnon-cooparon is evaluated as discussed in the main text.

\newpage
\subsection{\!\!\!\!\!\!\!Supplementary Note 4: NONLOCAL EXCITATION OF THE MAGNON SPIN CURRENT VIA A SUPERCONDUCTOR}


\label{nonlocal}

The system under consideration is depicted in Fig.~3 of the main text. It is assumed that the width $t$ of each of the FIs is smaller than the typical magnon wavelength in order to neglect inhomogeneities of the magnetization distribution in each of the FIs along the $x$-direction. From the other hand we assume $t \gg \xi_{\mathrm S}$. In this case we can consider each of the FIs as semi-infinite from the point of view of the superconductor. With these assumptions we solve the linearized Usadel equation in the superconductor with the following spatial profile of the exchange field:
\begin{eqnarray}
\mathbf{h}(x) = h_0(x)\mathbf{e}_z + \delta \mathbf{h}(x), \nonumber \\
h_0(x) = h_0 [\Theta(-\frac{d}{2}-x) + \Theta(x-\frac{d}{2})], \nonumber \\
\delta \mathbf{h}(x) = \delta \mathbf{h}_{\mathrm l} \Theta(-\frac{d}{2}-x) + \delta \mathbf{h}_{\mathrm r} \Theta(x-\frac{d}{2}),
\label{h_profile}
\end{eqnarray}
\begin{eqnarray}
\delta \mathbf{h}_{{\mathrm l},{\mathrm r}} \hat {\bm \sigma} = \delta h_{{\mathrm l},{\mathrm r}} {\mathrm e}^{-{\mathrm i}(\mathbf{k} \mathbf{r} + \omega t)\hat \sigma_z} \hat \sigma_x .
\label{h_profile2}
\end{eqnarray}
In order to catch the physical essence of the effect basing on the simplest equations, we again work in the linearized with respect to $\Delta$ and adiabatic approximation. Making use of the unitary transformation Eq.~(\ref{transform}) we come to the following equation for the dynamical triplet anomalous Green's function
\begin{eqnarray}
-{\mathrm i} \frac{D k^2}{4}[\delta \hat f_{\mathrm m}^{\mathrm R} - \hat \sigma_z \delta \hat f_{\mathrm m}^{\mathrm R} \hat \sigma_z] + {\mathrm i}\frac{D \partial_x^2 \delta \hat f_{\mathrm m}^{\mathrm R}}{2} = \varepsilon \delta \hat f_{\mathrm m}^{\mathrm R} - \frac{h_0}{2}\{\hat \sigma_z, \delta \hat f_{\mathrm m}^{\mathrm R} \} - \delta h(x)\hat \sigma_x f_{0,{\mathrm s}}^{\mathrm R} (x),
\label{f_spatial}
\end{eqnarray}
where $\delta h(x) = \delta h_{\mathrm l} \Theta(-\frac{d}{2}-x) + \delta h_{\mathrm r} \Theta(x-\frac{d}{2})$ and $f_{0,{\mathrm s}}^{\mathrm R} (x)$ is now also spatially inhomogeneous and should be found from the system of coupled equations:
\begin{eqnarray}
{\mathrm i}\frac{D \partial_x^2 f_{0,{\mathrm s}}^{\mathrm R}}{2} =
\varepsilon f_{0,{\mathrm s}}^{\mathrm R} - h_0(x)f_{0,z}^{\mathrm R}-\Delta, \nonumber \\
{\mathrm i}\frac{D \partial_x^2 f_{0,z}^{\mathrm R}}{2} =
\varepsilon f_{0,z}^{\mathrm R} - h_0(x)f_{0,{\mathrm s}}^{\mathrm R} .
\label{f0_spatial}
\end{eqnarray}
The singlet anomalous Green's function obeying these equations takes the form:
\begin{eqnarray}
f_{0,{\mathrm s}}^{\mathrm R} (x) = \left\{
\begin{array}{cc}
\frac{\Delta \varepsilon}{(\varepsilon+{\mathrm i} \delta)^2-h_0^2}+C_{+{\mathrm l}}{\mathrm e}^{\lambda_+(x+d/2)}+C_{-{\mathrm l}}{\mathrm e}^{\lambda_-(x+d/2)}, & x<-d/2, \\
\frac{\Delta_0}{\varepsilon}+C_1{\mathrm e}^{\lambda x} + C_2 {\mathrm e}^{-\lambda x}, & -d/2<x<d/2, \\
\frac{\Delta \varepsilon}{(\varepsilon+{\mathrm i} \delta)^2-h_0^2}+C_{+{\mathrm r}}{\mathrm e}^{-\lambda_+(x-d/2)}+C_{-{\mathrm r}}{\mathrm e}^{-\lambda_-(x-d/2)}, & x>d/2,
\end{array}\right.
\label{f_singlet_spatial}
\end{eqnarray}
where $\Delta_0$ and $\Delta$ are superconducting order parameters in the middle superconducting region $-d/2<x<d/2$ and in the left and right regions covered by the FIs, respectively. $\Delta<\Delta_0$ because of the order parameter suppression by the exchange field $h_0$. $\lambda = \sqrt{-2{\mathrm i}\varepsilon/D}$, $\lambda_{\pm} = \sqrt{-2{\mathrm i}(\varepsilon \pm h_0)/D}$. Constants $C_{\pm {\mathrm l},{\mathrm r}}$ take the form:
\begin{eqnarray}
C_{\pm {\mathrm l}} = C_{\pm {\mathrm r}} = \frac{1}{2}\left(\frac{\Delta}{\varepsilon \pm h_0}-\frac{\Delta_0}{\varepsilon} \right) \frac{\frac{\lambda_\pm}{\lambda}(1-\cosh \lambda d)-\sinh \lambda d}{(1+\frac{\lambda_\pm^2}{\lambda^2})\sinh \lambda d + 2 \frac{\lambda_\pm}{\lambda}\cosh \lambda d}.
\label{constants_0}
\end{eqnarray}
The solution of Eq.~(\ref{f_spatial}) takes the form $\hat f_{{\mathrm m}}^{\mathrm R} = f_{{\mathrm m},x} \sigma_x$ with
\begin{eqnarray}
\delta f_{\mathrm m}^{\mathrm R}(x) = \left\{
\begin{array}{cc}
C_{x{\mathrm l}}{\mathrm e}^{\lambda_k (x+d/2)} + \frac{\delta h_{\mathrm l} \Delta \varepsilon}{(\varepsilon+{\mathrm i}Dk^2/2)((\varepsilon+{\mathrm i} \delta)^2-h_0^2)} +~~~~~~\\ ~~~~~~\frac{2\delta h_{\mathrm l} C_{+ {\mathrm l}}{\mathrm e}^{\lambda_+ (x+d/2)}}{{\mathrm i}D(\lambda_k^2 - \lambda_+^2)}+\frac{2\delta h_{\mathrm l} C_{- {\mathrm l}}{\mathrm e}^{\lambda_- (x+d/2)}}{{\mathrm i}D(\lambda_k^2 - \lambda_-^2)}, & x<-d/2, \\
C_{x1}{\mathrm e}^{\lambda_k x} + C_{x2}{\mathrm e}^{-\lambda_k x}, & -d/2<x<d/2, \\
C_{x{\mathrm r}}{\mathrm e}^{-\lambda_k (x-d/2)} + \frac{\delta h_{\mathrm r} \Delta \varepsilon}{(\varepsilon+{\mathrm i}Dk^2/2)((\varepsilon+{\mathrm i} \delta)^2-h_0^2)} +~~~~~~\\ ~~~~~~\frac{2\delta h_{\mathrm r} C_{+ {\mathrm r}}{\mathrm e}^{-\lambda_+ (x-d/2)}}{{\mathrm i}D(\lambda_k^2 - \lambda_+^2)}+\frac{2\delta h_{\mathrm r} C_{- {\mathrm r}}{\mathrm e}^{-\lambda_- (x-d/2)}}{{\mathrm i}D(\lambda_k^2 - \lambda_-^2)}, & x>d/2,
\end{array}
\right. \nonumber
\end{eqnarray}
where $\lambda_k = \sqrt{-2{\mathrm i}(\varepsilon+{\mathrm i}Dk^2/2)/D}$ and constants take the form:
\begin{eqnarray}
C_{x{\mathrm l},{\mathrm r}} = -\frac{\delta h_{{\mathrm l},{\mathrm r}}\Delta \varepsilon}{2(\varepsilon+{\mathrm i}\frac{Dk^2}{2})[(\varepsilon+{\mathrm i}\delta)^2-h_0^2]}+
\frac{\delta h_{{\mathrm r},{\mathrm l}}\Delta \varepsilon {\mathrm e}^{-\lambda_kd}}{2(\varepsilon+{\mathrm i}\frac{Dk^2}{2})[(\varepsilon+{\mathrm i}\delta)^2-h_0^2]}-\nonumber \\
\frac{\delta h_{{\mathrm l},{\mathrm r}}}{{\mathrm i}D}\Bigl( \frac{1\mp\frac{\lambda_+}{\lambda_k}}{\lambda_k^2 - \lambda_+^2}C_{+ {\mathrm l},{\mathrm r}} + \frac{1\mp\frac{\lambda_-}{\lambda_k}}{\lambda_k^2 - \lambda_-^2}C_{- {\mathrm l},{\mathrm r}} \Bigr)+~~~~~~~~~~\nonumber\\
\frac{\delta h_{{\mathrm r},{\mathrm l}}{\mathrm e}^{-\lambda_k d}}{{\mathrm i}D}\Bigl( \frac{1\pm\frac{\lambda_+}{\lambda_k}}{\lambda_k^2 - \lambda_+^2}C_{+ {\mathrm r},{\mathrm l}} + \frac{1\pm\frac{\lambda_-}{\lambda_k}}{\lambda_k^2 - \lambda_-^2}C_{- {\mathrm r},{\mathrm l}} \Bigr).
\label{constants_x}
\end{eqnarray}
The spin polarization created by the triplet pairs is calculated from Eq.~(\ref{spin_current_2}). In the superconducting region under the left(right) FI it can be written as follows:
\begin{eqnarray}
    \bm s_{{\mathrm l},{\mathrm r}} = \bm s_0(x) + s_{{\mathrm {loc}}}(x)\frac{\delta \bm h_{{\mathrm l},{\mathrm r}}}{h_0} + s_{{\mathrm {nl}}}(x)\frac{\delta \bm h_{{\mathrm r},{\mathrm l}}}{h_0},
\label{s_spatial}
\end{eqnarray}
where the first term is the polarization induced by the equilibrium FI magnetization, $s_{{\mathrm {loc}}}$ is the polarization induced in the left (right) covered superconducting region by the magnon travelling in the left (right) FI and $s_{{\mathrm {nl}}}$ is the nonlocal part of the polarization induced by the magnon travelling in the right(left) covered superconducting region via the nonlocal triplet correlations penetrating from the other covered superconducting region. As it was mentioned above, we assume that the FI magnetization is homogeneous along the $x$-direction in each of the FIs. Consequently, only the averaged over the FI width value of the polarization enters the LLG equation. Therefore, we need to average Eq.~(\ref{s_spatial}) over the FI width in each of the superconducting regions under the FIs. Then $\overline{\bm s_0}$ and $\overline{s_{{\mathrm {loc}}}}$ contain terms of zero and first order with respect to the parameter $\xi_{\mathrm S}/t$ and we can neglect the terms $\sim \xi_{\mathrm S}/t \ll 1$ for simplicity without loss of important qualitative physics. At the same time $\overline{s_{{\mathrm {nl}}}}$ is $\sim \xi_{\mathrm S}/t$ because it is entirely determined by the correlations coming from the second FI region and decaying at $\xi_{\mathrm S}$. Under the described simplifying assumptions we obtain that $\mathbf{s}_0$ and $s_{{\mathrm {loc}}}\delta \mathbf{h}/h_0$ are described by the linearized versions of Eqs.~(\ref{spin_0}) and (\ref{spin_parallel}), respectively, and the nonlocal contribution to the polarization takes the form:
\begin{eqnarray}
\overline{s_{{\mathrm {nl}}}} = -N_{\mathrm F} \int d\varepsilon \tanh \frac{\varepsilon}{2T} {\rm Re}\Bigl[ \frac{h_0 \Delta \varepsilon {\mathrm e}^{-\lambda_k d}}{\lambda_k t [(\varepsilon + {\mathrm i} \delta)^2 - h_0^2]} \Bigl\{  \frac{\Delta \varepsilon}{2[(\varepsilon+{\mathrm i}\delta)^2-h_0^2](\varepsilon+{\mathrm i}Dk^2/2)}+ \nonumber \\
\frac{1}{{\mathrm i}D}\Bigl( \frac{C_{+,{\mathrm l}}}{\lambda_k(\lambda_k+\lambda_+)}+\frac{C_{-,{\mathrm l}}}{\lambda_k(\lambda_k+\lambda_-)} \Bigr)\Bigr\} \Bigr].~~~~
\label{s_nonlocal}
\end{eqnarray}
Taking into account the torque resulting from the nonlocal polarization, we obtain the coupled system of the LLG equations for the both FIs:
\begin{eqnarray}
\frac{\partial \mathbf{m}_{{\mathrm {l,r}}}}{\partial t} = - \gamma \mathbf{m}_{{\mathrm {l,r}}} \times \mathbf{H}_
{{\mathrm {eff}}} + \alpha \mathbf{m}_{{\mathrm {l,r}}} \times \frac{\partial \mathbf{m}_{{\mathrm {l,r}}}}{\partial t} + \frac{J}{d_{{\mathrm {FI}}}}\overline{s_{{\mathrm {nl}}}} \mathbf{m}_{{\mathrm {l,r}}} \times \mathbf{m}_{{\mathrm {r,l}}}.
\label{LLG_coupled}
\end{eqnarray}
The last term couples the both magnetizations. Then we express $\mathbf{m}_{{\mathrm {l,r}}} = \mathbf{m}_0 + \delta \mathbf{m}_{{\mathrm {l,r}}}$ as a sum of the equilibrium magnetization $\mathbf{m}_0$ and the magnon contribution $\delta \mathbf{m}_{{\mathrm {l,r}}}$.  Linearizing Eqs.~(\ref{LLG_coupled}) with respect to $\delta \mathbf{m}_{{\mathrm {l,r}}}$ and solving them for the magnon dispersion, we obtain:
\begin{eqnarray}
\omega_\pm (k) = \omega_0 + \tilde D_{\mathrm m} k^2 \mp \frac{J}{d_{{\mathrm {FI}}}}\overline{s_{{\mathrm {nl}}}},
\label{dispersion}
\end{eqnarray}
where $\omega_0 +\tilde D_{\mathrm m} k^2 $ is the magnon dispersion in a separate FI/S bilayer with the renormalized stiffness $\tilde D_{\mathrm m} = D_{\mathrm m} +\delta D_{\mathrm m}$ and the last term accounts for the hybridization and splitting of the magnon modes due to the interaction via the superconductor. The eigenvectors of the linearized with respect to $\delta \mathbf{m}_{{\mathrm {l,r}}}$ version of Eqs.~(\ref{LLG_coupled}) corresponding to the eigenfrequencies Eq.~(\ref{dispersion}) take the form:
\begin{eqnarray}
\left(
\begin{array}{c}
\delta \mathbf{m}_{\mathrm l,\pm} \\
\delta \mathbf{m}_{\mathrm r,\pm}
\end{array}
\right) =
\left(
\begin{array}{c}
1 \\
\pm 1
\end{array}
\right),
\label{eigenvectors}
\end{eqnarray}
that is, the eigenmodes of the system are represented by the symmetric and antisymmetric combination of the uncoupled magnons. The value of the frequency splitting $\Delta \omega = (J/d_{\mathrm {FI}})\overline{s_{\mathrm {nl}}}$ can be estimated as $\Delta \omega \sim 10^9 \exp(-d/\xi_{\mathrm S}) $ Hz for the material and geometric parameters used above and $t/\xi_{\mathrm S} \sim 10$.

The coupling between the uniform modes of two ferromagnetic insulators via a superconductor layer has recently been investigated~\cite{Ojajarviarxiv}. However, the nonzero-wavevector excitations considered here allow for a realization of magnon directional coupler based on a fundamentally different physical principle than it has been proposed earlier~\cite{Wang2018,Sadovnikov2019,Wang2020}. Indeed magnons of a given frequency injected into the coupled region have different wave vectors $k_\pm \approx k_0 \pm \Delta k$, where $\Delta k = k_0\Delta \omega/\tilde D_{\mathrm m} k_0^2$. The coupling length admitted by this design is given in Eq.~(12) of the main text. Among the other advantages of this coupling principle we can mention more compactness of the proposed setup, because the strength of the dipole-dipole coupling is strongly reduced with lowering the thickness $d_{{\mathrm {FI}}}$ of the FI layers along the $z$-direction~\cite{Wang2018}, while in the framework of the proposed mechanism the coupling strength is $\propto d_{{\mathrm {FI}}}^{-1}$, that is  ultra-thin ferromagnetic films are more favorable. Second, the superconducting coupling can be switched on/off by any means, which are known to control superconductivity: temperature, magnetic field, voltage. An interesting perspective is to investigate the possibility to control the coupling strength via the superconducting phase. Our proposed design also enables an analogous coupler for magnons in antiferromagnetic lines, because the dynamical triplets should also be generated there, but the stray fields are weak and, therefore, the dipole-dipole coupling principle does not work well.

\section{References}
\bibliography{MagCoop}

\end{document}